TOC Figure:

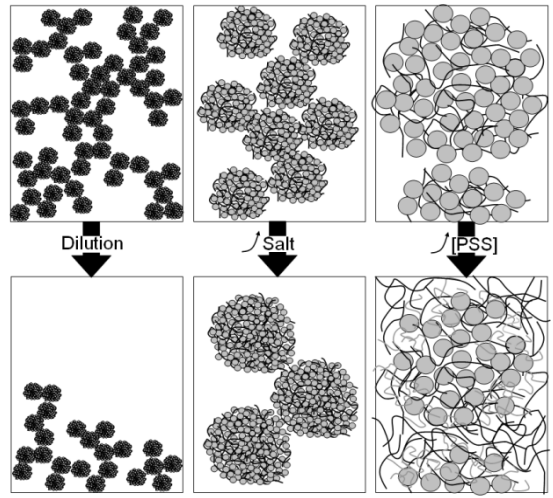

# Multiple Scale Reorganization of Electrostatic Complexes of PolyStyrene Sulfonate and Lysozyme

*Fabrice Cousin[1,*], Jérémie Gummel[1], Daniel Clemens[2], Isabelle Grillo[3], and François Boué[1]*

[1]Laboratoire Léon Brillouin, CEA Saclay 91191 Gif sur Yvette, Cedex France

[2] Helmholtz-Zentrum Berlin für Materialien und Energie GmbH, Glienicker Straße 100, 14109 Berlin-Wannsee Germany

[3]Institut Laue-Langevin, 6 rue Jules Horowitz, B.P. 156, 38042 Grenoble Cedex 9, France

[*] Corresponding author: fabrice.cousin@cea.fr

**Abstract**

We report on a SANS investigation into the potential for these structural reorganization of complexes composed of lysozyme and small PSS chains of opposite charge if the physicochemical conditions of the solutions are changed after their formation. Mixtures of solutions of lysozyme and PSS with high matter content and with an introduced charge ratio [-]/[+]$_{intro}$ close to the electrostatic stoichiometry, lead to suspensions that are macroscopically stable. They are composed at local scale of dense globular primary complexes of radius ~ 100 Å; at a higher scale they are organized fractally with a dimension 2.1. We first show that the dilution of the solution of complexes, all other physicochemical parameters remaining constant, induces a macroscopic destabilization of the solutions but does not modify the structure of the complexes at submicronic scales. This suggests that the colloidal stability of the complexes can be explained by the interlocking of the fractal aggregates in a network at high concentration: dilution does not break the local aggregate structure but it does destroy the network. We show, secondly, that the addition of salt does not change the almost frozen inner structure of the cores of the primary complexes, although it does encourage growth of the complexes; these coalesce into larger complexes as salt has partially screened the electrostatic repulsions between two primary complexes. These larger primary complexes remain aggregated with a fractal dimension of 2.1. Thirdly, we show that the addition of PSS chains up to $[-]/[+]_{intro} \sim 20$, after the formation of the primary complex with a $[-]/[+]_{intro}$ close to 1, only slightly changes the inner structure of the primary complexes. Moreover, in contrast to the synthesis achieved in the one-step mixing procedure where the proteins are unfolded for a range of $[-]/[+]_{intro}$, the native conformation of the proteins is preserved inside the frozen core.

# 1 Introduction

Soft matter systems generally display a multi-scale structure; they are composed of structures which can vary greatly with space scale. Depending on the spatial scale, these structures may be either at thermodynamic equilibrium or out of equilibrium. It is thus possible for a system to be at equilibrium at local scales but not at larger scales, since characteristically large objects have very long rearrangement times. The rearrangement process may thus involve various dynamics over a wide range of times. Thus slow spontaneous rearrangements can occur in systems; if their interesting properties are controlled by their structures at large scales (colloidal stability, for example) the impact on these properties can be dramatic. On the other hand, since soft matter systems are extremely sensitive to "low fields", i. e. small changes in various external parameters (from PH, temperature to mechanical constraint or electromagnetic fields) can trigger reorganization processes; they could be efficiently used to tune and improve the properties of the system. Reorganization processes induced by small changes can also fruitfully be used to understand how a given system has reached and maintained its final structure.

These features are found of course in various colloidal systems, particularly when they are complex - i.e. involving more than merely particle size and inter-particle distance, and, therefore, in polyelectrolyte-protein complexes of opposite charge. These latter are now attracting growing interest [2, 3, 4], especially if they present protein-polysaccharide complexes [5], given the potential application of these systems for the food or pharmaceutical industries (drug release [6], biochips [7], fractionation [8], stabilization of emulsions [9], etc). Moreover, the design of new architectures involving polyelectrolyte chains (such as polyelectrolyte multilayers [10] or polyelectrolyte dendrimers [11] enabling the immobilization of proteins on colloidal particles [12, 13]) enlarges the potential scope of these systems.

In spite of the apparent complexity of polyelectrolyte-protein complexes (the interactions depend on parameters such as the pH, ionic strength, concentration, charge density of the polyelectrolyte, charge distribution on the protein, rigidity of the chain, hydrophobicity of the molecules), these systems usually share similar features: they form a hierarchy of structures at different scales, with an initial structure on local scales composed of dense complexes with a globular shape of a few hundred Å; these will be referred to below as primary complexes. These primary complexes can be observed experimentally [14, 15, 16, 17, 18, 19, 20, 21] and by simulation [22, 23]. At higher scales (> 1000 Å) they tend to phase separate - either into two fluid phases (complex coacervation) [24], or into a lower turbid phase, coexisting with a much clearer supernatant phase and containing fractal aggregates sufficiently large to sediment, and in some cases finally precipitate [25, 26].

The generality of this structure can be explained by the fact that the system is essentially governed by one type of physical interaction, electrostatics, acting on different spatial scales. At local scale the critical conditions required for complexation at room temperature appear closely linked to ionic strength and surface charge density: the adsorption/desorption limit is an inverse function of the Debye length $\kappa^{-1}$ [27]. When the ionic strength becomes too high, the screening of the electrostatic interactions prevents the formation of complexes [28, 29]. The presence of heterogeneous charges on the protein surface enhances complexation [15, 16, 30, 31]. When complexation occurs, aggregation is maximum when the charges brought to solution by the two species are at stoichiometry [15, 16, 17, 18, 19, 20, 22]. The stoichiometry is recovered inside the core of a complex [19]. For highly charged systems, complexation is endothermic [18] and thus entropically driven due to the release of condensed counterions, as proposed in [32, 33] and experimentally checked in [34]. At higher scales it has recently been shown that $\kappa$ tunes the size of the primary complexes [20]. Coacervation is usually observed with poorly charged systems and precipitation with highly charged systems. Other interactions, such as H-bond or hydrophobic interactions, can play a role in certain systems [2, 35]. Chain stiffness can also tune electrostatic interactions, since the binding of spherical macroions on polyelectrolytes decreases as chain stiffness rises [28, 29, 36].

This spatial hierarchical organization enables us to identify certain simple key parameters which are instrumental in the process of simple reorganization by external stimuli. This is rich in potential for further applications, such as the salt-induced release of lipase from polyelectrolyte complex micelles recently described by Linkhoud et al [37]. There are certain properties, however, perhaps of relevance to these applications (e.g. the stability of the colloids or activity of the proteins), that can be altered during reorganization. If we are to optimize the use of proteins/polyelectrolyte complexes in applications, it is therefore important, to obtain a detailed description of the mechanisms involved, in terms of both the physicochemical parameters and length scales. This is the question we address in this paper.

We focus on complexes made of lysozyme and polystyrene sulfonate (PSS) of opposite charge. There is a significant amount of information available from previous studies for this system. We know that when PSS chains are in a semi-dilute regime after interaction with proteins [38] they form a gel crosslinked by proteins [14]. When in dilute regime however, they form dense globular primary complexes of radius ~ 100 Å with a neutral core from the electrostatic point of view and high compactness (~ 30% of matter) [19, 20]; they are organized at a higher scale in fractal fashion with a dimension 2.1 by reaction-limited cluster aggregation (RLCA) [25]. At very high PSS content, the protein native shape can be unfolded [14, 39]; this unfolding is specific to PSS and occurs in various PSS architectures (i.e. brushes or stars) [40, 41].

The system in the dilute regime (globules) appears particularly suitable for the study of the reorganization process:

- First, the structural characteristics of the globules can be fixed completely and in a reproducible way by the initial mixing synthesis.

- Secondly, whereas primary globules are formed on time scales smaller than a second, the final state of the system is out of equilibrium [14] and some reorganization can occur over much longer time scales (several days or weeks).

- Thirdly, the scales for the interactions within and between the globular primary complexes are distinct, with different strengths and ranges [20]. This increases the diversity of the reorganization processes.

- Fourthly, the specific unfolding of lysozyme by PSS can also be used to characterize the permeation of PSS within the primary complexes and the subsequent reorganization induced by such permeation. We will specifically follow lysozyme conformation, which is of particular interest because it is one of the parameters of that determines enzymatic activity of the complexes.

We investigate the possibility of reorganising the complexes at the 3 scales relevant to the system (Figure 1):

- at large scales (> 1000 Å), we investigate the organization between the aggregates of primary complexes. This scale governs the macroscopic properties of the suspension such as its colloidal stability or viscosity.

  - at intermediate scales (100 Å – 1000 Å) we look at the outer characteristics of the primary complexes (size, charge, specific area); these are fundamental in the control of properties such as enzymatic activity.

  - at local scales (10 Å – 100 Å), we examine the inner organization of primary complexes for information on protein conformation, the direct interactions between components, compactness of globules, etc.

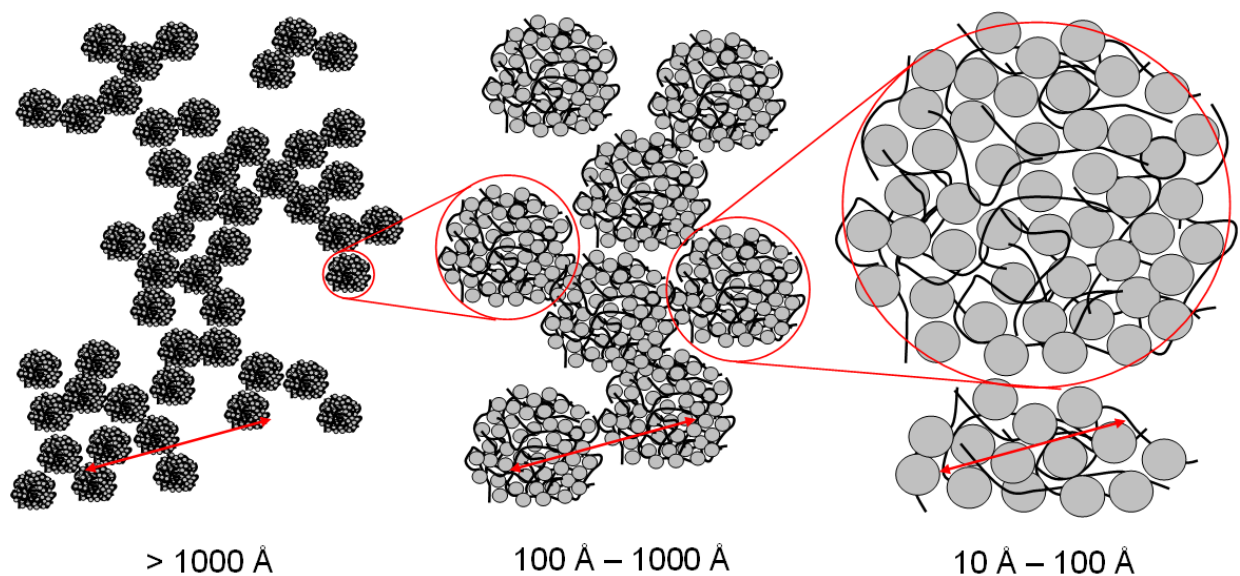

Figure 1: Sketch of the PSS-lysozyme system at the 3 different scales examined in the present paper.

To follow the structural changes we took advantage of small-angle neutron scattering (SANS), which is capable of probing the 3 different scales. SANS is expanding in popularity as a technique for determining the local structure of protein-polyelectrolyte complexes, within coacervates in particular. It has been successfully applied recently to BSA-PSS mixtures [42], pectin-PSS mixtures [21], β-lactoglobulin/pectin mixtures [43], agar-gelatine mixtures [44], BSA-chitosan [45] and BSA-PDAMAC [46]. We present the influence of three specific parameters on the potential for these reorganization processes within the solution of PSS/lysozyme complexes: (i) dilution of the solution; (ii) addition of salt to the solution; (iii) addition of a large amount of PSS chains to the solution. We show that each parameter enabled us to tune the reorganization within the complexes at one of the three scales being considered.

## 2 Materials and methods

*2.1 Sample preparation*

The sulfonation of the polystyrene chains was performed in-house using the method in [47], itself derived from the Makowski method [48]. We used here fully sulfonated PS chains, *i.e* each PSS repetition unit bears a negative charge, and has a molal volume 108 g/mole. We used deuterated polystyrene chains of 50 repetitions units with very low polydispersity ($M_w/M_n \sim 1.03$), purchased from Polymer Standard Service.

The lysozyme (molal mass 14298, charge +11 at pH 4.7) was purchased from Sigma and used without further purification.

The samples were all prepared in two steps. They were first treated, as described in our former papers [14, 19, 20], as follows. They were prepared at pH 4.7 in an acetic acid/acetate buffer ($CH_3COOH/CH_3COO^-$). The counterions were $Na^+$. The buffer concentration was set by default to $5 \; 10^{-2}$ mol/L. Two solutions, one of lysozyme and one of PSSNa, were prepared separately in the buffer at a concentration two times higher than the final concentration of the sample. They were then mixed and slightly shaken to be homogenized. Apart from the blank samples designed specifically for comparison with the ones to which PSS is added on pre-formed complexes (see below), we used a lysozyme concentration of 40g/L to get a good SANS signal. The PSSNa concentration was then adapted to produce samples with charge ratios introduced in solution (denoted $[-]/[+]_{intro}$) varying from 0.6 to 3.33, according to the experiment. The charge ratio takes into account the structural charges and not the effective charge. It is thus calculated with the net charge of lysozyme (+11 at pH 4.7; [+] = 0.03 M for 40 g/L) and with one negative charge per sulfonated PS monomer on the PSS chains. As soon as the mixture had been prepared, a very turbid liquid was obtained in all cases. These ranges of lysozyme concentration and $[-]/[+]_{intro}$ produced samples which were macroscopically homogeneous and showed no significant change over the following days or even months (see the state of the system diagram in [14]).

After the formation of complexes, all the samples were left alone for at least two days, before starting on the second step in the procedure.

The second step depends on the parameter that was tested to change the interactions in the system:

- for the dilution experiment at constant ionic strength after the formation of complexes, we chose to work with a sample with $[-]/[+]_{intro} = 0.66$. The sample was split into 3 samples of 1 ml. Two of these were diluted with an acetate buffer at $I = 5\ 10^{-2}$ mol/L, one by a factor 3.33 and the other by 10. The third was kept as a reference sample.

- for the experiment at increased ionic strength after the formation of complexes, we chose to use samples with $[-]/[+]_{intro} = 1.66$ in the acetate buffer at $I = 5\ 10^{-2}$ mol/L. The salinity of the buffer was increased with NaCl to $5\ 10^{-1}$ mol/L. Starting from $I = 5\ 10^{-2}$ mol/L in a 2 mL sample, 200 μL of a NaCl solution at either 0.6, 1.7 or 5 mol/L were added two days after the preparation of the solution of complexes, to reach ionic strengths of up to $I = 5\ 10^{-1}$ mol/L. The additional volume of the NaCl solution was sufficiently low to consider the concentration of complexes as unchanged. After the addition of salt the samples remained liquid and turbid but visually homogeneous.

- for the experiment in which PSS chains were added after the formation of complexes, we started with a series of 4 samples with $[-]/[+]_{intro} = 3.33$. A concentrated solution of PSS was then added to the samples 2 days after their initial preparation, to values of 8, 13 and 20 for $[-]/[+]_{intro}$ respectively.

For the sake of comparison, we also prepared 3 samples in a one-step procedure with the following $[-]/[+]_{intro}$: 8 ([lyso] = 25 g/L; [PSS] = 0.15 mol/L), 13 ([lyso] = 20 g/L; [PSS] = 0.2 mol/L), 20 ([lyso] = 20 g/L; [PSS] = 0.3 mol/L).

*2.2 SANS experiments*

The SANS measurements were performed on either the HMI's V4 spectrometer (HMI, Berlin, Germany) in a q-range lying between $3.10^{-3}$ and $3.10^{-1}$ Å$^{-1}$ or on the ILL's D11 spectrometer (ILL, Grenoble, France) in a q-range lying between $6.5\ 10^{-4}$ and $3.3\ 10^{-1}$ Å$^{-1}$. All the measurements were performed at atmospheric pressure and room temperature.

In order to obtain the PSSNa signal and lysozyme signal independently, each PSS/protein composition was measured in two different solvents: once in a fully D$_2$O buffer matching the neutron scattering length density of deuterated PSSNa, and once in a 57%/43% H$_2$O/D$_2$O mixture matching the neutron scattering length density of lysozyme.

In order to obtain the scattered intensities on an absolute scale, standard corrections were applied for sample volume, neutron beam transmission, empty cell signal subtraction, detector efficiency, subtraction of incoherent scattering and solvent buffer.

# 3 SANS results and discussion: Evolution of the structures of the complexes when the interactions are changed after their formation

## 3.1 Colloidal stability of the complexes after dilution (large length scale)

In our previous studies of the structure of dense globular complexes formed with small chains [19, 20, 25] performed at a content of 40g/L, the samples generally displayed macroscopic homogeneity on timescales of weeks. We have shown, however, in [14] that there are some regions of the state diagram at lower protein content where a turbid fraction, decanted at the bottom of the cell, coexists with a clear supernatant. In some parts of this biphasic region of the state diagram, the PSS chains are in a dilute regime after interaction with lysozyme; we would thus expect the structure of the complexes to be made of fractal aggregates of dense globules of ~ 10 nm of radius. This suggests that the macroscopic stability of the system of globules can be explained simply by the "scaffolding" effect obtained when some of the branches of the partly interpenetrated fractal aggregates interlock. The volume fraction occupied by the primary aggregates must be superior to the overlapping threshold $\Phi_{agg}^* = N_{agg} V_{globule} / (R_{agg})^3 \sim N_{agg}^{(1-3/D_f)}$, where $N_{agg}$ is the mean aggregation number. The $N_{agg}$ and $R_{agg}$ are unfortunately too large to be measured by either SANS (see [25]) or electron microscopy (for which only a very broad aggregate size distribution can be observed).

In order to test our hypothesis we took a sample macroscopically stable on a timescale of weeks and diluted it by factors of 3.33 and 10, as described in Materials and Methods above. Unlike the undiluted sample, the diluted samples phase separated within a few minutes into a turbid precipitate and a clear supernatant. The supernatant-to-precipitate volume ratio was about 1:3 for the sample diluted by the factor 3.33, and 3:4 for the sample diluted by the factor 10. We then measured the neutron scattering at small angles of lysozyme for the three samples, in a solvent matching the PSS scattering length density. As the measurements were to be taken while the diluted samples decanted, they were performed at the ILL, to benefit from the high neutron flux down to q as low as $6.10^{-4}$ Å$^{-1}$. Before measuring at each of the 4 configurations used, the sample was gently shaken in order to ensure that the solution was macroscopically homogeneous. The sample was placed in the neutron beam and the scattering recorded for 3 minutes. This allowed sufficient time for good data collection before excessive sedimentation. The merging of I(q) on the absolute scale between the 4 configurations was always perfect (see Figure 2). This proves that the 3-minute sedimentation took place in the same fashion all four times, and also that the structure of the sample at the scale investigated by the neutrons was similar whatever the number of sedimentation/gentle shaking cycles. In other words, sedimentation does not induce further aggregation of the complexes at these scales.

The scattering curves of the 3 samples are shown in Figure 2. They all display exactly the same features, similar to those described in our previous papers [14, 19, 20, 25]: a correlation peak at 0.2 Å$^{-1}$ corresponding to two lysozymes in contact inside the globular primary complexes, a $q^{-4}$ behaviour at intermediate q corresponding to the surface scattering of the globular primary complexes, and a $q^{-2.1}$ behaviour at low q corresponding to their fractal organization. The fitting of the scattering in both the intermediate and low q ranges gives us the mean size of the globules $R_{mean}$. The method is extensively described in ref [19] and recalled in Supporting Information. For the undiluted sample, we obtained $R_{mean}$ = 100 Å as in previous results [20]. For the two diluted samples, we obtained the same curve, although shifted in intensity. This means that the three samples share similar structure up to $2\pi/0.0006$ Å$^{-1}$ ~ 10 000 Å, which was the largest scale in direct space probed in our scattering experiments. This gives us a lower limit for the aggregation number of primary complexes

$N_{agg}$ [19] which is higher than 140 for the $R_{mean}$ obtained here. The measured scattering intensity reduced, after dilution by a factor 3, by a dividing factor of approximately 8. This reduction ratio, after dilution by a factor 10, is equal to ~ 26. The scattering reduction ratios are thus different from the dilution ratios. This is due to the sedimentation of the samples which occurs during measurement: dilution reduces the effective volume fraction of complexes illuminated by the neutron beam, compared to the real volume fraction. (Please note that the degree of sedimentation after 3 minutes remains the same, since 3/10 ~ 8/26: the sedimentation rate does not change with dilution).

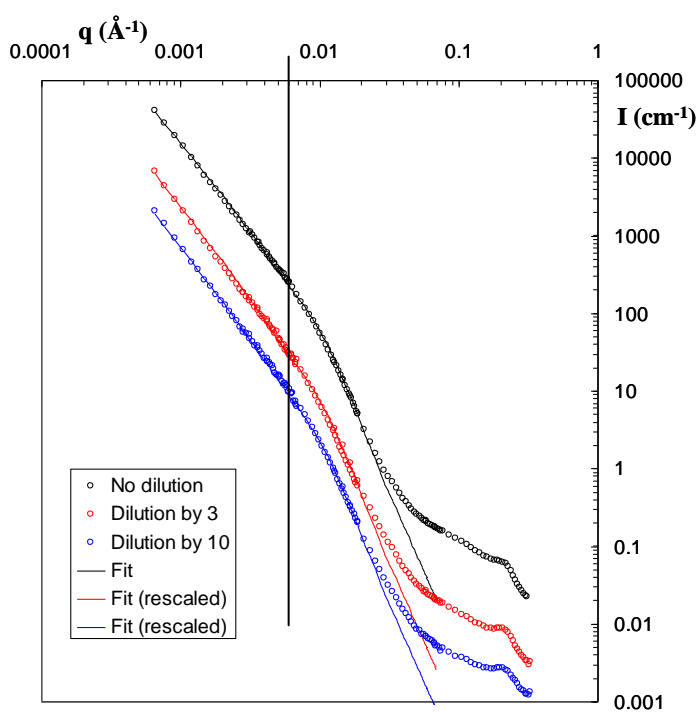

**Figure 2:** SANS lysozyme scattering of samples obtained in a 100% $D_2O$ acetate buffer at pH 4.7. Ionic strength I = 5 $10^{-2}$ mol/L, $[-]/[+]_{intro}$ = 0.66 and $[Lyso]_{intro}$ = 40g/L, after synthesis, after dilution in the same buffer by a factor 3 and by a factor 10. All curves are in absolute scale. The full lines correspond to the fits.

This simple shift between curves proves that the structural reorganization leading to sedimentation occurs only at very large scales, in direct space. The $R_{mean}$ of the globules does not change; neither does the fractal dimension of the aggregates of globules, which stays at 2.1. This strongly suggests that the sedimentation provoked by dilution is only due to the de-scaffolding of the fractal aggregates of primary complexes: the macroscopic stability of the suspension of aggregates is governed by volume fraction $\Phi_{agg}$: they are stable if $\Phi_{agg}$ is higher than the overlapping threshold $\Phi_{agg}^*$. Assuming all PSS ([PSS] = 0.02M, volume fraction 2.16 cc/L) is in the aggregates, and using the internal fraction given in Table 1 of S.I. we obtain for the undiluted mixture, $\Phi_{agg} \sim 0.4$. The simplistic expression of $\Phi_{agg}^* \sim N_{agg}^{(1-3/Df)} \cdot \phi_{inner}$, can be used only very crudely because we do not have access to the size distribution of the fractal aggregates. From $N_{agg} > 140$ given above, we get $\Phi_{agg}^* \sim 0.2$. This gives an order of magnitude which agrees with observation, but is extremely approximate.

**3.2 Influence of the addition of salt on the primary complexes (intermediate length scale)**

We now present the evolution in the structures of the primary complexes when the ionic strength of the buffer was increased from $I = 5 \cdot 10^{-2}$ mol/L to $I = 5 \cdot 10^{-1}$ mol/L after the first preparatory step; the aim is to examine the effect of screening the electrostatic interactions in the system. We measured both lysozyme scattering and PSS scattering, independently, using SANS. Both series of spectra displayed features typical of the scattering of curves of primary complexes (Figure SI.1 in the Supporting Information – see below) aggregated in a fractal way (see part 3.1). As shown in [19] and recalled in supplementary information, the fit of the scattering of both lysozyme and PSS gives the mean radius of the primary globular complexes $R_{mean}$, the mean number of lysozymes per complex $N_{lyso\_comp}$, the size of the outer PSS shell (corona) when it exists, the volume fraction of organic matter within the complexes $\Phi_{inner}$ and the inner charge ratio within the complexes $[-]/[+]_{inner}$. Figure 3 and Table SI.1 show the results of the fits for these four quantities, while the scattering and its fits are plotted in both Figure 4 and Figure SI.1. We comment on these below.

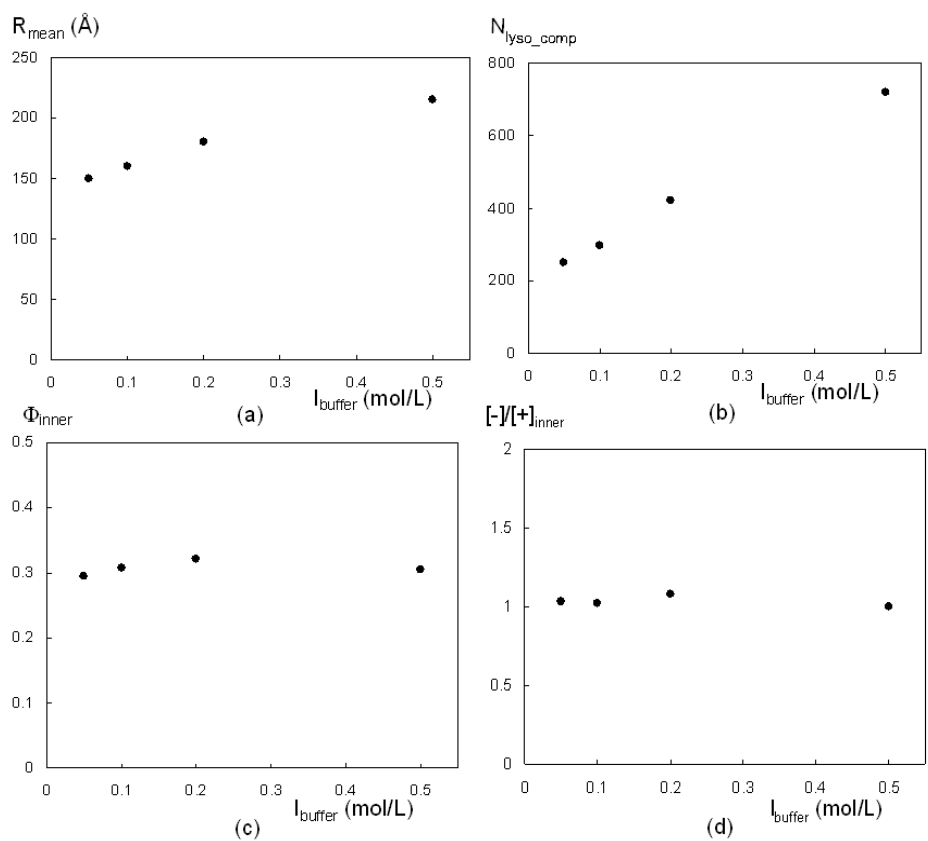

**Figure 3:** Overview of the effect of additional salt (increase of ionic strength $I_{buffer}$), set after synthesis step 1 for samples at pH 4.7, and $[-]/[+]_{intro}$ = 1.66 and $[Lyso]_{intro}$ = 40g/L: (a) $R_{mean}$ versus $I_{buffer}$; (b) $N_{lyso\_comp}$ versus $I_{buffer}$; (c) $\Phi_{inner}$ versus $I_{buffer}$; (d) $[-]/[+]_{inner}$ versus $I_{buffer}$.

The first striking result, visible in Figure 3c, concerns the compactness of the cores of the primary complexes $\Phi_{inner}$. They definitely do not vary when the salinity of the solution is increased. Moreover (Figure 3d), the cores stays neutral: $[-]/[+]_{inner}$ retains its value of ~ 1 whatever the ionic strength, while $[-]/[+]_{intro}$ is 1.66. This confirms that the cores of primary complexes do not undergo change; we can consider them "frozen". The freezing of the core comes here from the very high content of matter in the inner of the primary complexes. In less dense systems such as coavervates of whey protein and arabic Gum [49], some slow reorganization can occur with time as it has been shown in this system that the whey proteins diffuse 10 times faster than the Arabic gum within the coacervates [50]. **The reorganization of solutions of primary complexes induced by modifying the electrostatic interactions does not occur within the primary complexes.**

**On larger scales, however, the size of the primary complexes is strongly affected** by the ionic strength of the buffer; this can be seen in Figure 3a (giving $R_{mean}$), and it is clear in Figure 4. We present here the lysozyme scattering in a Porod representation $I(q)q^4 = f(q)$. In this representation, the maximum observed at low q is linked to the value of the globule radius [19]. It is shifted towards low q when salt is added, which means that the globule radius increases. In addition, a plateau can be observed at intermediate q in this view; it is well known that its height is linear to the specific area S/V of the complexes, inversely proportional to the characteristic size of the globules ( $\sim 3/R_{mean}$). This specific area decreases if salt is added, because $R_{mean}$ increases.

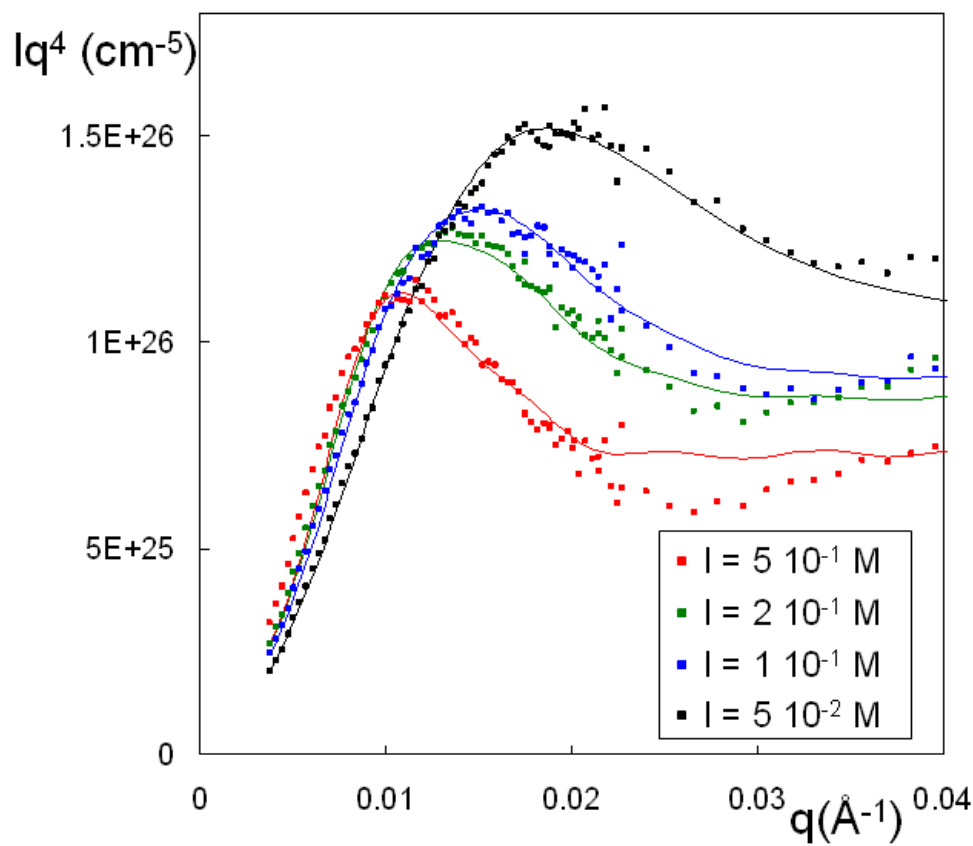

**Figure 4:** samples at pH 4.7, and $[-]/[+]_{intro}$ = 1.66 and $[Lyso]_{intro}$ = 40g/L with different I of the buffer set after the formation of the complexes. Porod representation of the lysozyme scattering at low q. The full lines correspond to the fits.

The last quantity measured, $N_{lyso\_comp}$, increases from 250 at I = 5 $10^{-2}$ mol/L to 720 at 5 $10^{-1}$ mol/L (Figure 3 b). This is in accordance with the increase size of the primary complexes with a constant compactness and a constant $[-]/[+]_{inner}$. The mean number of PSS repetitions units per complex also increases in order to keep $[-]/[+]_{inner}$ constant, as also shown in Table SI.1. However, as the electrostatic stoichiometry is kept in the solution, the number of free PSS chains in solution, in excess from an electrostatic point of view, stays constant. The ionic strength therefore has a direct impact on the interaction of the primary complexes; it tunes both $R_{mean}$ and $N_{lyso\_comp}$.

This behaviour mirrors that observed when the ionic strength of the buffer was changed in the initial solution (one-step synthesis): the higher the salinity, the larger the primary complexes [20]. However the size of these primary complexes is slightly smaller when salt is added after their formation (for example we see $R_{mean}$ > 300 Å at I = 5 $10^{-1}$ mol/L with a one-step synthesis, and we see $R_{mean}$ ~ 250 Å at the same I when salt is added after synthesis). In particular, if we translate the growth of $N_{lyso\_comp}$ as a function of I (Figure 3) into growth as a function of $\kappa$ ($\kappa \sim I^{1/2}$ being the inverse of the Debye length noted $\kappa^{-1}$), we find it to be rather slower than the exponential growth obtained in [20] for the same system. In other words, the final size of the primary complexes is slightly dependent on the history of salinity. We can nevertheless conclude that the reorganization with extra salt only occurs at the scale of the size of the primary complexes and is essentially controlled by the Debye length of the solution. This is coherent with the process we described in [20], to explain why the growth of the primary complexes stops at a finite size: during growth the complexes progressively repel each other by electrostatic repulsion as their charge increases. As this repulsion dominates, the complexes stop growing and behave like charged colloids; this results in a Reaction Limited Cluster Aggregation (RLCA) process, and thus to an aggregate fractal dimension Df of 2.1. If the electrostatic interactions are screened by the salt, the potential between the complexes reverts below the critical value for the shift in the aggregation process; reorganization becomes possible. Since the internal composition of the primary complexes remains unchanged, this resembles the **coalescence** of small primary globules into larger globules as salinity increases, as is the case with emulsions droplets. What occurs between neighbouring globules is a "soft rearrangement": in the simplest scenario, the fractal network is not forced to "break and reform" during the process. The connections between the aggregates rearrange locally allowing progressive merging, but do not completely destroy the scaffolding.

The induced reorganization does not change the volume fraction of complexes in solution because the compactness of complexes stays constant, but could change the mean aggregation number $N_{agg}$ of complexes due to the reorganization of the scaffolding. Such reorganization could have induced a macroscopic destabilization of the suspension, like the one induced by dilution in part 3.1. However, in the range of salinity considered here, the change of $N_{agg}$ remains too low to cross the overlapping threshold $\Phi_{agg}*$ because the suspensions remain macroscopically homogeneous whatever the change of salinity, as pointed out in the Materials and Methods section.

**3.3 Access for PSS chains to the core of the primary complexes (short length scale)**

We showed in the previous section that the reorganization within the solutions of complexes following the addition of a large amount of salt occurs on scales larger than those of primary globular complexes, and that there is no internal structural reorganization of the core of the primary complexes since compactness and $[-]/[+]_{intro}$ remain constant. We proposed that the core of the complexes is "frozen" after their formation through electrostatic attractions. In order to test this assumption, we have devised a special test to examine the potential accessibility of the core of the globules once they are formed. We exploit a specificity of the lysozyme/PSS system - the unfolding of lysozyme by PSS chains. This occurs at high $[-]/[+]_{intro}$, and is attributed to the interactions between the hydrophobic patches of lysozyme and the hydrophobic backbone of PSS after initial strong electrostatic complexation [14]. This unfolding, clearly evidenced by a change in the scattering of lysozyme, leads to a transition from turbid to limpid samples. For the small chains considered here, it occurs at $[-]/[+]_{intro}$ = 10. The principle of our experiment is to synthesize complexes at low $[-]/[+]_{intro}$ (3.33, see Materials and Methods) and to add *a posteriori* PSS chains to the samples, in order to reach high $[-]/[+]_{intro}$ and check whether unfolding also occurs, like in the one-step mixing. A concentrated PSS solution was thus added to the reference solution at $[-]/[+]_{intro}$ = 3.33 two days after its synthesis, to reach values of 8, 13 and 20 for $[-]/[+]_{intro}$. SANS measurements were performed twice on such samples, to check if the system was close to equilibrium: the scattering was first recorded 2 days after the addition of PSS chains and then again 9 days after adding the PSS chains. The results of the two measurements were strictly identical. This proves than there are no slow reorganizations within the primary complexes, in accordance with the fact that the inner core is almost frozen. One of the two sets of results is plotted in Figure 5. It can be compared with the scattering obtained with the reference samples (prepared with the one-step procedure).

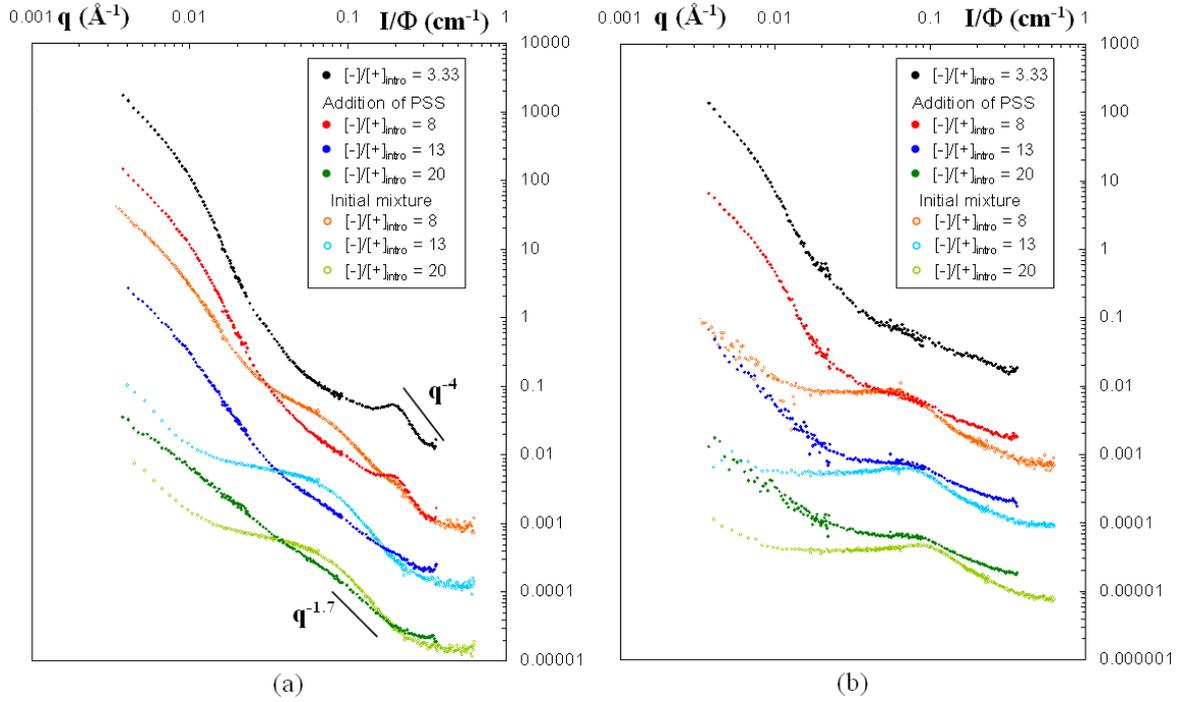

**Figure 5:** Comparison of SANS data for samples with $[-]/[+]_{intro}$ lying between 3.33 and 20 (pH 4.7, I = 5 $10^{-2}$ ), with all PSS chains in the initial solution or added two days after the formation of the complexes. (a) Lysozyme scattering: all curves are shifted from one to another by a decade, for clarity. The intensity in absolute scale corresponds to the $[-]/[+]_{intro}$ = 3.33 sample. (b) PSS chains scattering: all curves are shifted from one to another by a decade for clarity. The intensity in absolute scale corresponds to the $[-]/[+]_{intro}$ = 3.33 sample.

The scattering at $[-]/[+]_{intro}$ = 3.33 presents the typical features of primary complexes described in the previous sections. In particular, at large q, the lysozyme scattering decays like $q^{-4}$ due to the globular native shape of lysozyme. For larger ratios, let us now describe the one-step case: as soon as $[-]/[+]_{intro}$ = 8, at large q, the peak 0.2 Å$^{-1}$ due to the contact between two proteins and the $q^{-4}$ scattering become less visible; the low q scattering is abated. These changes are complete at [-]/[+]$_{intro}$ = 13 and evolve no further at [-]/[+]$_{intro}$ = 20. The change in behaviour from a $q^{-4}$ to a $q^{-1.7}$ is due to the unfolding of proteins, which then scatter in a self-avoiding walk in a fashion similar to excluded-volume polymer chains [14, 51]. The peak at 0.2 Å$^{-1}$ thus also disappears. A shoulder appears above 0.1 Å$^{-1}$; it is reminiscent of the "polyelectrolyte peak" in the PSS scattering (see below) which we attribute to the fact that the unfolded proteins chains now decorate the transient network of PSS chains. Since the unfolding of proteins induces the destruction of the globules and their aggregates, the scattering at low q falls by two orders of magnitudes.

If we now look at the scattering of [-]/[+]$_{intro}$ = 8, it becomes clear that it has features of the scattering of globules (strong low q scattering) coexisting with unfolded proteins (a shoulder below around 0.1 Å$^{-1}$). This reflects the fact that the [-]/[+]$_{intro}$ is close to the [-]/[+]$_{intro}$ value of the macroscopic limpid/turbid transition.

Let us now look at the scattering after the second step of mixing: the scattering of the [-]/[+]$_{intro}$ = 8 sample is unchanged (globular complexes) compared with the [-]/[+]$_{intro}$ = 3.33 sample. The scattering for [-]/[+]$_{intro}$ = 13 shows changes, but weaker than in the one-step mixing at the same [-]/[+]$_{intro}$: they are only comparable to those observed for [-]/[+]$_{intro}$ = 8 for one-step mixing. Even for [-]/[+]$_{intro}$ = 20, while changes are more important, the unfolding is not complete: there remains a small q scattering, indicating that some complexes with unfolded proteins are still present in the solution. It is clear that two to three times as many PSS chains, compared with one-step mixing, are necessary to unfold lysozyme. **In summary, when PSS is added after the formation of primary complexes, behaviour similar to that observed after only one-step mixing is obtained, but the transition between folded and unfolded proteins is shifted towards much higher [-]/[+]$_{intro}$.**

The PSS scattering confirms these results. We know that when proteins are native, PSS chains are shrunk [14]: their persistence length $L_p$ is reduced by the bridging interactions with lysozyme. Consequently, the abscissa $q^*$ of the "polyelectrolyte peak" (due to a privileged distance among PSS chains) is shifted to lower q. If lysozyme is unfolded, the chain shrinking vanishes and q* becomes equal to that in pure solutions. We can follow these effects quite simply after the first step of mixing: for $[-]/[+]_{intro}$ = 13, a "polyelectrolyte peak" becomes clearly visible, and the low q scattering disappears: lysozyme is unfolded. For $[-]/[+]_{intro}$ = 20, there is no more changes which shows that the process is completed at $[-]/[+]_{intro}$ = 13 . For $[-]/[+]_{intro}$ = 8, the polyelectrolyte peak is visible but significant globule scattering remains at low q: lysozyme is not completely unfolded. Therefore, in the one-step mixing, the crossover from one regime of $[-]/[+]_{intro}$ ratios to the other is simultaneous for PSS scattering and lysozyme scattering. This simultaneity is also seen after the second step of mixing: for the low q scattering, no change at $[-]/[+]_{intro}$ = 8, but the beginning of a decrease for $[-]/[+]_{intro}$ = 13, which continues at $[-]/[+]_{intro}$ = 20. For the "polyelectrolyte peak" no sign at $[-]/[+]_{intro}$ = 8, a shoulder at $[-]/[+]_{intro}$ = 13, and a soft maximum at $[-]/[+]_{intro}$ = 20. The q* at $[-]/[+]_{intro}$ = 20 however has a lower value after two-step mixing than one-step mixing. This is understandable, given that the PSS chains are not all involved in the polyelectrolyte semidilute solution: when PSS chains are added *a posteriori,* a fraction of their total number remains embedded in the primary complexes. The semidilute solution is thus less concentrated.

This proves that the inner of the core of the primary complexes is weakly accessible. The permeation of the added chains inside the primary complexes is very limited. This is consistent with the high compactness of the primary complexes. Their core is almost all frozen as soon as they are formed. When adding even more PSS chains, up to very large amounts, the lysozyme unfolds at last, but with slow kinetics. This permeation at high content of PSS chains is probably induced by the high osmotic pressure of the PSS chains outside the primary complexes.

This experiment shows that it is possible to exploit the freezing of the core to produce complexes that shelter proteins in their native state (hence keeping their enzymatic activity), although it has been established that the binding of proteins to PSS polyelectrolyte brushes denatures them [41].

**4 Conclusions and perspectives**

We have shown that it is possible to induce structural reorganization within solutions of PSS/lysozyme complexes after their formation and that the typical space scale of the reorganization process depends on the physico-chemical parameter modified (see Figure 6):

(i) A simple dilution does not change the structure of the fractal aggregates of primary complexes of radius ~ 100 Å at scales lower than ~ 10000 Å; it can however lead to macroscopic destabilization of the solutions, because the concentration passes below the overlapping threshold of the fractal aggregates of the globules, which breaks down the scaffolding they were forming.

(ii) The addition of salt only changes the structure of complexes at scale equal or higher than the radius of the primary complexes: the screening of the electrostatic interactions between complexes enables the fusion of several complexes into larger ones, but there are no change in the inner structure of the core, which keeps its neutral electric charge and constant compactness

(iii) There is almost no structural reorganization at scales lower than those of the primary complexes: reorganization is triggered neither by the addition of salt (even large quantities) nor by the addition of PSS chains, even up to concentrations that unfold lysozyme after one-step mixing. The core appears as almost "frozen", and the proteins inside stay native.

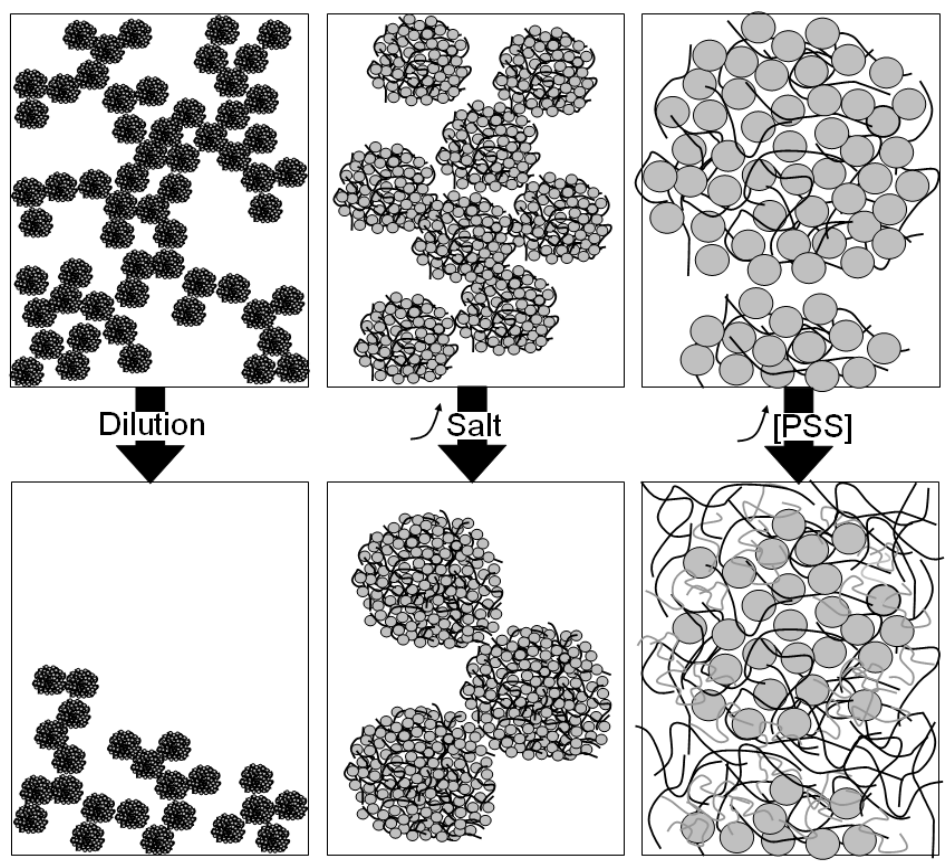

**Figure 6:** Representations of the 3 kinds of reorganization occurring in the system as a function of the scale investigated.

Let us finally note that the dependency of the reorganization processes on both the physicochemical parameters and length scale is important when considering the application of such complexes, in the development of carriers for drug delivery, for example. The structural reorganization of the complexes in response to changes in the physicochemical parameters is indeed now known and controlled. Consider the example of enzyme delivery: we have shown that ionic strength acts in PSS/lysozyme complexes as a controller of their specific area, and thus of he active enzymes exposed to the solvent, without leading to subsequent unfolding unlike in the case of proteins hosted in a PSS brush. Alternatively, the fact that the core of primary complexes is "frozen" enables it to shelter proteins in their native state while they are exposed to environments that should in principle denature them.

**Supporting Information:** The SANS spectra for complexes to which salt has been added in solution after the formation of complexes and their corresponding fits as well as summary of the fitting procedure of the SANS data are available free of charge via the internet on: http://pubs.acs.org.

**Acknowledgements:** The authors thank Robert Corner for his careful reading of the manuscript.